\documentclass[12pt,a4paper,twocolumn]{article}
\usepackage[utf8]{inputenc}
\usepackage[T1]{fontenc}
\usepackage{amsmath}
\usepackage{amsfonts}
\usepackage{amssymb}
\usepackage{fullpage}
\usepackage{bbold}
\usepackage{xcolor}
\usepackage{natbib}
\usepackage[font={small}]{caption}
\usepackage[affil-it]{authblk}

\usepackage{longtable}
\usepackage{float}
\usepackage[hmargin={0.6in,0.6in},vmargin={1in,1in}]{geometry}
\usepackage[space]{grffile}
\usepackage{subcaption}
\usepackage{graphicx}
\usepackage{epstopdf} 
\usepackage{algpseudocode}
\usepackage{booktabs}
\usepackage{abstract}

\author[1,2,3]{Fran\c{c}ois Lafond}

\author[4]{Aimee Gotway Bailey}
\author[5]{Jan David Bakker}
\author[2]{Dylan Rebois}
\author[6,1,7]{Rubina Zadourian}
\author[2,8,9,10]{Patrick McSharry}
\author[1,6,11,12]{J. Doyne Farmer}

\affil[1]{\small Institute for New Economic Thinking at the Oxford Martin School
}
\affil[2]{Smith School for Enterprise and the Environment, University of Oxford
}
\affil[3]{London Institute for Mathematical Sciences
}
\affil[4]{
U.S. Department of Energy
}
\affil[5]{Department of Economics, University of Oxford
}

\affil[6]{Mathematical Institute, University of Oxford
}

\affil[7]{Max Planck Institute for the Physics of Complex Systems,
Germany
}

\affil[8]{Carnegie Mellon University Africa,
Rwanda
}
\affil[9]{African Center of Excellence in Data Science, University of Rwanda
}

\affil[10]{Oxford Man Institute of Quantitative Finance, University of Oxford
}

\affil[11]{Computer Science Department, University of Oxford
}
\affil[12]{Santa Fe Institute
 USA}

\newcommand{\possessivecite}[1]{\citeauthor{#1}'s \citeyearpar{#1}}

\title{How well do experience curves predict technological progress?  A method for making distributional forecasts\footnote{Acknowledgements: This project was primarily supported by the European Commission project {FP7-ICT-2013-611272} (GROWTHCOM) and by Partners for a New Economy. We also gratefully acknowledge support from the European Commission project {H2020-730427} (COP21 RIPPLES) and the Institute for New Economic Thinking. We are grateful to Giorgio Triulzi and two anonymous referees for comments on an earlier draft. Contact: francois.lafond@inet.ox.ac.uk, doyne.farmer@inet.ox.ac.uk.}
}
\date{\today}

\begin{document}

\twocolumn[
 \begin{@twocolumnfalse}
    \maketitle
\begin{abstract}
Experience curves are widely used to predict the cost benefits of increasing the deployment of a technology.  But how good are such forecasts?  Can one predict their accuracy a priori?  In this paper we answer these questions by developing a method to make distributional forecasts for experience curves.  We test our method using a dataset with proxies for cost and experience for 51 products and technologies and show that it works reasonably well.  The framework that we develop helps clarify why the experience curve method often gives similar results to simply assuming that costs decrease exponentially.  To illustrate our method we make a distributional forecast for prices of solar photovoltaic modules.



\end{abstract}
\vspace{1cm}
  \end{@twocolumnfalse}
]
\saythanks

\section{Introduction}
\label{section:intro}

Since \possessivecite{wright1936factors} study of airplanes, it has been observed that for many products and technologies the unit cost of production tends to decrease by a constant factor every time cumulative production doubles \citep{thompson2012relationship}. This relationship, also called the experience or learning curve, has been studied in many domains.\footnote{
See \citet{yelle1979learning,dutton1984treating,anzanello2011learning} and for energy technologies \citet{neij1997use,isoard2001technical,nemet2006beyond,kahouli2009testing,junginger2010technological,candelise2013dynamics}.}
 It is often argued that it can be useful for forecasting and planning the deployment of a particular technology \citep{ayres1969technological,sahal1979theory,martino1993technological}. However in practice experience curves are typically used to make point forecasts, neglecting prediction uncertainty.  Our central result in this paper is a method for making {\it distributional forecasts}, that explicitly take prediction uncertainty into account.  We use historical data to test this and demonstrate that the method works reasonably well.

Forecasts with experience curves are usually made by regressing historical costs on cumulative production. In this paper we recast the experience curve as a time series model expressed in first-differences: the change in costs is determined by the change in experience. We derive a formula for how the accuracy of prediction varies as a function of the time horizon for the forecast, the number of data points the forecast is based on, and the volatility of the time series.  We are thus able to make distributional rather than point forecasts.  Our approach builds on earlier work by \citet{farmer2016how} that showed how to do this for univariate forecasting based on a generalization of Moore's law (the autocorrelated geometric random walk with drift).  Here we apply our new method based on experience curves to solar photovoltaics modules (PV) and compare to the univariate model.

Other than \citet{farmer2016how}, the two closest papers to our contribution here are \citet{alberth2008forecasting} and \citet{nagy2013statistical}. Both papers tested the forecast accuracy of the experience curve model, and performed comparisons with the time trend model. \citet{alberth2008forecasting} performed forecast evaluation by keeping some of the available data for comparing forecasts with actual realized values.\footnote{
\citet{alberth2008forecasting} produced forecasts for a number (1,2, $\dots$ 6) of doublings of cumulative production. Here instead we use time series methods so it is more natural to compute everything in terms of calendar forecast horizon.}
Here, we build on the methodology developed by \citet{nagy2013statistical} and \citet{farmer2016how}, which consists in performing systematic hindcasting.  That is, we use an estimation window of a constant (small) size and perform as many forecasts as possible.
As in \citet{alberth2008forecasting} and \citet{nagy2013statistical}, we use several datasets and we pool forecast errors to construct a distribution of forecast errors. We think that out-of-sample forecasts are indeed good tests for models that aim at predicting technological progress. However, when a forecast error is observed, it is generally not clear whether it is ``large'' or ``small'', from a statistical point of view. And it is not clear that it makes sense to aggregate forecast errors from technologies that are more or less volatile and have high or low learning rates. 

A distinctive feature of our work is that we actually calculate the expected forecast errors. As in \citet{farmer2016how}, we derive an approximate formula for the theoretical variance of the forecast errors, so that forecast errors from different technologies can be normalized, and thus aggregated in a theoretically grounded way. As a result, we can check whether our empirical forecast errors are in line with the model. We show how in our model forecast errors depend on future random shocks, but also parameter uncertainty, as is only seldomly acknowledged in the literature (for exceptions, see \citet{vigil1994estimating} and \citet{van2008introducing}).

\citet{alberth2008forecasting} and \citet{nagy2013statistical} compared the forecasts from the experience curve, which we call Wright's law, with those from a simple univariate time series model of exponential progress, which we call Moore's law. While \citet{alberth2008forecasting} found that the experience curve model was vastly superior to an exogenous time trend, our results (and method and dataset) are closer to the findings of \citet{nagy2013statistical}: univariate and experience curves models tend to perform similarly, due to the fact that for many products cumulative experience grows exponentially. When this is the case, we cannot expect experience curves to perform much better than an exogenous time trend unless cumulative experience is very volatile, as we explain in detail here.

We should emphasize that this comparison is difficult because the forecasts are conditioned on different variables:  Moore's law is conditioned on time, while Wright's law is conditioned on experience.  Which of these is more useful depends on the context.  As we demonstrate, Moore's law is more convenient and just as good for business as usual forecasts for a given time in the future.  However, providing there is a causal relationship from experience to cost, Wright's law makes it possible to forecast for policy purposes \citep{way2017wright}.

Finally, we depart from \citet{alberth2008forecasting}, \citet{nagy2013statistical} and most of the literature by using a different statistical model. As we explain in the next section, we have chosen to estimate a model in which the variables are first-differenced, instead of kept in levels as is usually done. From a theoretical point of view, we believe that it is reasonable to think that the stationary relationship is between the increase of experience and technological progress, instead of between a level of experience and a level of technology.
In addition, we will also introduce a moving average noise, as in \citet{farmer2016how}. This is meant to capture some of the complex autocorrelation patterns present in the data in a parsimonious way, and increase theoretical forecast errors so that they match the empirical forecast errors.

Our focus is on understanding the forecast errors from a simple experience curve model\footnote{
We limit ourselves to showing that the forecast errors are compatible with our model being correct, and we do not try to show that they could be compatible with the experience curve model being spurious.}.
The experience curve, like any model, is only an approximation.  Its simplicity is both a virtue and a detriment.  The virtue is that the model is so simple that its parameters can usually be estimated well enough to have predictive value based on the short data sets that are typically available\footnote{
For short data sets such as most of those used here, fitting more than one parameter often results in degradation in out-of-sample performance \citep{nagy2013statistical}.}.
The detriment is that such a simple model neglects many effects that are likely to be important. A large literature starting with \citet{arrow1962economic} has convincingly argued that learning-by-doing occurs during the production (or investment) process, leading to decreasing unit costs. But innovation is a complex process relying on a variety of interacting factors such as economies of scale, input prices, R\&D and patents, knowledge depreciation effects, or other effects captured by exogenous time trends.\footnote{
For examples of papers discussing these effects within the experience curves framework, see \citet{argote1990persistence}, \citet{berndt1991practice}, \citet{isoard2001technical}, \citet{papineau2006economic}, \citet{soderholm2007empirical}, \citet{jamasb2007technical}, \citet{kahouli2009testing}, \citet{bettencourt2013determinants}, \citet{benson2015quantitative} and \citet{nordhaus2014perils}. }
For instance, \citet{candelise2013dynamics} argue that there is a lot of variation around the experience curve trend in solar PV, due to a number of unmodelled mechanisms linked to industrial dynamics and international trade, and \citet{sinclair2000s} argued that the relationship between costs and experience is due to experience driving expectations of future production and thus incentives to invest in R\&D. Besides, some have argued that simple exponential time trends are more reliable than experience curves. For instance \citet{funk2015rapid} noted that significant technological improvements can take place even though production experience did not really accumulate, and   \citet{magee2016quantitative} found that in domains where experience (measured as annual patent output) did not grow exponentially, costs still had a exponentially decreasing pattern, breaking down the experience curve. Finally, another important aspect that we do not address is reverse causality \citep{kahouli2009testing,nordhaus2014perils,witajewski2015bending}: if demand is elastic, a decrease in price should lead to an increase in production. Here we have intentionally focused on the simplest case in order to develop the method.

\section{Empirical framework}

\subsection{The basic model}
\label{section:basicmodel}

Experience curves postulate that unit costs decrease by a constant factor for every doubling of cumulative production\footnote{
For other parametric models relating experience to costs see \citet{goldberg2003statistical} and \citet{anzanello2011learning}.}.
This implies a linear relationship between the log of the cost, which we denote $y$, and the log of cumulative production which we denote $x$:
\begin{equation}
y_t= y_0 + \omega x_t.
\label{eq:Wrightlawdet}
\end{equation}
This relationship has also often been called ``the learning curve'' or the experience curve. We will often call it ``Wright's law'' in reference to Wright's original study, and to express our agnostic view regarding the causal mechanism. Generally, experience curves are estimated as
\begin{equation}
y_t= y_0 + \omega x_t + \iota_t,
\label{eq:Wrightlawlevel}
\end{equation}
where $\iota_t$ is i.i.d. noise. However, it has sometimes been noticed that residuals may be autocorrelated. For instance \citet{womer1983estimation} noticed that autocorrelation ``seems to be an important problem'' and \citet{lieberman1984learning} ``corrected'' for autocorrelation using the Cochrane-Orcutt procedure.\footnote{See also \citet{mcdonald1987new}, \citet{hall1985experience}, and \citet{goldberg2003statistical} for further discussion of the effect of autocorrelation on different estimation techniques.} \citet{bailey2012forecasting} proposed to estimate Eq.\eqref{eq:Wrightlawdet} in first difference 
\begin{equation}
y_t - y_{t-1}= \omega (x_t - x_{t-1})+ \eta_t,
\label{eq:Wrightlaw}
\end{equation}
where $\eta_t$ are i.i.d errors $\eta_t \sim \mathcal{N}(0,\sigma_\eta^2)$. In Eq.\eqref{eq:Wrightlaw}, noise accumulates so that in the long run the variables in level can deviate significantly from the deterministic relationship. To see this, note that (assuming $x_0=\log(1)=0$) Eq. \eqref{eq:Wrightlaw} can be rewritten as
\[
y_t= y_0 + \omega x_t + \sum_{i=1}^t \eta_i,
\]
which is the same as Eq.\eqref{eq:Wrightlawlevel} except that the noise is accumulated across the entire time series.  In contrast, Eq.\eqref{eq:Wrightlawlevel} implies that even in the long run the two variables should be close to their deterministic relationship. 

If $y$ and $x$ are I(1)\footnote{A variable is I(1) or integrated of order one if its first difference $y_{t+1}-y_t$ is stationary.}, Eq.\eqref{eq:Wrightlawlevel} defines a cointegrated relationship. We have not tested for cointegration rigorously, mostly because unit root and cointegration tests have uncertain properties in small samples, and our time series are typically short and they are all of different length. Nevertheless, we have run some analyses suggesting that the difference model may be more appropriate. First of all in about half the cases we found that model \eqref{eq:Wrightlawlevel} resulted in a Durbin-Watson statistic lower than the $R^2$, indicating a risk of spurious regression and suggesting that first-differencing may be appropriate\footnote{
Note, however, that since we do not include an intercept in the difference model and since the volatility of experience is low, first differencing is not a good solution to the spurious regression problem.}.
Second, the variance of the residuals of the level model was generally higher, so that the tests proposed in \citet{harvey1980comparing} generally favored the first-difference model. Third, we ran cointegration tests in the form of Augmented Dickey-Fuller tests on the residuals of the regression \eqref{eq:Wrightlawlevel}, again generally suggesting a lack of cointegration. While a lengthy study using different tests and paying attention to differing sample sizes would shed more light on this issue, in this paper we will use Eq. \eqref{eq:Wrightlaw} (with autocorrelated noise). The simplicity of this specification is also motivated by the fact that we want to have the same model for all technologies, we want to be able to calculate the variance of the forecast errors, and we want to estimate parameters with very short estimation windows so as to obtain as many forecast errors as possible.

We will compare our forecasts using Wright's law with those of a univariate time series model which we call Moore's law 
\begin{equation}
y_{t}-y_{t-1}=\mu+n_t.
\label{eq:Moorelaw}
\end{equation}
This is a random walk with drift. The forecast errors have been analyzed for i.i.d. normal $n_t$ and for $n_t=v_t+\theta v_{t-1}$ with i.i.d. normal $v_t$ (keeping the simplest forecasting rule) in \citet{farmer2016how}.
As we will note throughout the paper, if cumulative production grows at a constant logarithmic rate of growth, i.e.  $x_{t+1}-x_t=r$ for all $t$, Moore's and Wright's laws are observationally equivalent in the sense that Eq. \eqref{eq:Wrightlaw} becomes Eq. \eqref{eq:Moorelaw} with $\mu=\omega r$. This equivalence has already been noted by \citet{sahal1979theory} and \citet{ferioli2009learning} for the deterministic case. \citet{nagy2013statistical}, using a dataset very close to ours, showed that using trend stationary models to estimate the three parameters independently (Eq. \eqref{eq:Wrightlawlevel} and regressions of the (log) costs and experience levels on a time trend), the identity $\hat{\mu}=\hat{\omega} \hat{r}$ holds very well for most technologies. Here we will replicate this result using difference stationary models.

\subsection{Hindcasting and surrogate data procedures}
\label{section:hindcastingandsurrogate}

To evaluate the predictive ability of the models, we follow closely \citet{farmer2016how} by using hindcasting to compute as many forecast errors as possible and using a surrogate data procedure to test their statistical compatibility with our models. Pretending to be in the past, we make pseudo forecasts of values that we are able to observe and compute the errors of our forecasts. More precisely, our procedure is as follows. We consider all periods for which we have ($m+1$) years of observations (i.e. $m$ year-to-year growth rates) to estimate the parameters, and at least one year ahead to make a forecast (unless otherwise noted we choose $m$=5). For each of these periods, we estimate the parameters and make all the forecasts for which we can compute forecast errors. Because of our focus on testing the method and comparing with univariate forecasts, throughout the paper we assume that cumulative production is known in advance. 
Having obtained a set of forecast errors, we compute a number of indicators, such as the distribution of the forecast errors or the mean squared forecast error, and compare the empirical values to what we expect given the size and structure of our dataset. 

To know what we expect to find, we use an analytical approach as well as a surrogate data procedure. The analytical approach simply consists in deriving an approximation of the distribution of forecast errors. However, the hindcasting procedure generates forecast errors which, for a single technology, are not independent\footnote{For a review of forecasting ability tests and a discussion of how the estimation scheme affects the forecast errors, see \citet{west2006forecast} and \citet{clark2013advances}.}. However, in this paper we have many short time series so that the problem is somewhat limited (see Appendix \ref{appendix:checktheo}). Nevertheless, we deal with it by using a surrogate data procedure: we simulate many datasets similar to ours and perform the same analysis, thereby determining the sampling distribution of any statistics of interest.

\subsection{Parameter estimation}

To simplify notation a bit, let $Y_t=y_t - y_{t-1}$ and $X_t=x_t - x_{t-1}$ be the changes of $y$ and $x$ in period $t$.  We estimate Wright's exponent from Eq.\eqref{eq:Wrightlaw} by running an OLS regression through the origin. Assuming that we have data for times $i=1...(m+1)$, minimizing the squared errors gives\footnote{
Throughout the paper, we will use the hat symbol for estimated parameters when the estimation is made using only the $m+1$ years of data on which the forecasts are based.  When we provide full sample estimates we use the tilde symbol.}
\begin{equation}
\hat{\omega}=\frac{\sum_{i=2}^{m+1} X_{i} Y_{i} }{\sum_{i=2}^{m+1} X_{i}^2}.
\label{eq:omegahat}
\end{equation}
Substituting $\omega X_i+\eta_i$ for $Y_i$, we have
\begin{equation}
\hat{\omega}=\omega+\frac{\sum_{i=2}^{m+1} X_{i} \eta_{i} }{\sum_{i=2}^{m+1} X_{i}^2}.
\label{eq:omegahatbias}
\end{equation}
The variance of the noise $\sigma_\eta^2$ is estimated as the regression standard error
\begin{equation}
\hat \sigma_\eta^2=\frac{1}{m-1}\sum_{i=2}^{m+1}(Y_{i}-\hat \omega X_{i})^{2}.
\label{eq:sigmahat}
\end{equation}

For comparison, parameter estimation in the univariate model Eq. \eqref{eq:Moorelaw} as done in \citet{farmer2016how} yields the sample mean $\hat{\mu}$ and variance $\hat{K}^2$ of $Y_t \sim \mathcal{N}(\mu, K^2)$.

\subsection{Forecast errors}
\label{section:forecasterrors}

Let us first recall that for the univariate model Eq. \eqref{eq:Moorelaw}, the variance of the forecast errors is given by \citep{sampson1991effect,clements2001forecasting,farmer2016how}
\begin{equation}
E[\mathcal{E}_{M,\tau}^2]= K^2 \left(\tau+\frac{\tau^2}{m} \right),
\label{eq:MSFEmoore}
\end{equation}
where $\tau$ is the forecast horizon and the subscript $_M$ indicates forecast errors obtained using ``Moore'''s model. It shows that in the simplest model, the expected squared forecast error grows due to future noise accumulating ($\tau$) and to estimation error ($\tau^2/m$). These terms will reappear later so we will use a shorthand
\begin{equation}
A \equiv \tau+\frac{\tau^2}{m},
\label{eq:A}
\end{equation}

We now compute the variance of the forecast errors for Wright's model. If we are at time $t=m+1$ and look $\tau$ steps ahead into the future, we know that
\begin{equation}
y_{t+\tau}=y_t+\omega (x_{t+\tau}-x_t)+\sum_{i=t+1}^{t+\tau} \eta_{i}.
\label{eq:yttau}
\end{equation}
To make the forecasts we assume that the future values of $x$ are known, i.e. we are forecasting costs conditional on a given growth of future experience.  This is a common practice in the literature \citep{meese1983empirical,alberth2008forecasting}. More formally, the point forecast at horizon $\tau$ is
\begin{equation}
\hat{y}_{t+\tau}=y_t+\hat \omega (x_{t+\tau}-x_t).
\label{eq:yhatttau}
\end{equation}
The forecast error is the difference between Eqs. \eqref{eq:yttau} and \eqref{eq:yhatttau}, that is
\begin{equation}
\mathcal{E}_\tau \equiv y_{t+\tau}-\hat{y}_{t+\tau}=(\omega-\hat \omega) \sum_{i=t+1}^{t+\tau} X_i+\sum_{i=t+1}^{t+\tau} \eta_{i}.
\end{equation}

We can derive the expected squared error. Since the $X_i$s are known constants, using $\hat \omega$ from Eq. \eqref{eq:omegahatbias} and the notation $m+1=t$, we find
\begin{equation}
E[\mathcal{E}_\tau^2]= \sigma_\eta^2 \left(\tau + \frac{\left(\sum_{i=t+1}^{t+\tau}  X_{i}\right)^2}{\sum_{i=2}^{t}X_i^2} \right).
\label{eq:MSFE}
\end{equation}

\subsection{Comparison of Wright's law and Moore's law}

\citet{sahal1979theory} was the first to point out that in the deterministic limit the combination of exponentially increasing cumulative production and exponentially decreasing costs gives Wright's law.  Here we generalize this result in the presence of noise and show how variability in the production process affects this relationship.

Under the assumption that experience growth rates are constant ($X_i=r$) and letting $m=t-1$, Eq. (\ref{eq:MSFE}) gives the result that the variance of Wright's law forecast errors are precisely the same as the variance of Moore's law forecast errors given in Eq.\eqref{eq:MSFEmoore}, with $\hat{K}=\hat{\sigma}_\eta$.  To see how the fluctuations in the growth rate of experience impact forecast errors we can rewrite Eq. (\ref{eq:MSFE}) as
\begin{equation}
E[\mathcal{E}_\tau^2]= \sigma_\eta^2 \left(\tau +\frac{\tau^2}{m}  \hspace{2mm} \frac{\hat{r}_{(f)}^2}{\hat{\sigma}^2_{x,(p)} + \hat{r}_{(p)}^2} \right),
\label{eq:MSFErewritten}
\end{equation}
where $\hat{\sigma}^2_{x,(p)}$ refers to the estimated variance of past experience growth rates, $\hat{r}_{(p)}$ to the estimated mean of past experience growth rates, and $\hat{r}_{(f)}$ to the estimated mean of future experience growth rates.\footnote{
The past refers to data at times ($1,\dots, t$) and the future to times $(t+1, \dots, t+\tau)$.}  

This makes it clear that the higher the volatility of experience ($\sigma_x^2$), the lower the forecast errors. This comes from a simple, well\--known fact of regression analysis: high variance of the regressor makes the estimates of the slope more precise. Here the high standard errors in the estimation of $\omega$ (due to low $\sigma_x^2$) decrease the part of the forecast error variance due to parameter estimation, which is associated with the term $\tau^2/m$.

This result shows that, assuming Wright's law is correct, for Wright's law forecasts to work well (and in particular to outperform Moore's law), it is better to have cumulative production growth rates that fluctuate a great deal.  Unfortunately for our data this is typically not the case. Instead, empirically cumulative production follows a fairly smooth exponential trend. To explain this finding we calculated the stochastic properties of cumulative production assuming that production is a geometric random walk with drift $g$ and volatility $\sigma_q$. In Appendix \ref{appendix:sigmaX}, using a saddle point approximation for the long time limit we find that $E[X]\equiv r \approx g$ and
\begin{equation}
\mbox{Var}[X] \equiv \sigma_x^2 \approx \sigma_q^2 \tanh (g/2),
\label{eq:sigmaX}
\end{equation}
where $\tanh$ is the hyperbolic tangent function. We have tested this remarkably simple relationship using artificially generated data and we find that it works reasonably well.
  
These results show that cumulative production grows at the same rate as production.  More importantly, since $0<\tanh(g/2)<1$ (and here we assume $g > 0$), the volatility of cumulative production is lower than the volatility of production. This is not surprising:  it is well-known that integration acts as a low pass filter, in this case making cumulative production smoother than production.  Thus if production follows a geometric random walk with drift, experience is a smoothed version, making it hard to distinguish from an exogenous exponential trend. When this happens Wright's law and Moore's law yield similar predictions. This theoretical result is relevant to our case, as can be seen in the time series of production and experience plotted in Fig. \ref{fig:prod} and Fig. \ref{fig:expe}, and the low volatility of experience compared to production reported in Table \ref{table:parameterestimates} below.

\subsection{Autocorrelation}
\label{section:forecasterrors-Autocorrelation}

We now turn to an extension of the basic model. As we will see, the data shows some evidence of autocorrelation. Following \citet{farmer2016how}, we augment the model to allow for first order moving average autocorrelation.  For the autocorrelated Moore's law model ( ``Integrated Moving Average of order 1'')
\[
y_{t}-y_{t-1}=\mu+v_t+\theta v_{t-1},
\]
\citet{farmer2016how} obtained a formula for the forecast error variance when the forecasts are performed assuming no autocorrelation
\begin{equation}
E[\mathcal{E}_{M,\tau}^2]=\sigma_v^2 \Big[-2\theta + \Big(1+\frac{2(m-1)\theta}{m} +\theta^2\Big) A \Big],
\label{eq:MSFEMT}
\end{equation}
where $A=\tau+\frac{\tau^2}{m}$ (Eq.\eqref{eq:A}). Here we extend this result to the autocorrelated Wright's law model
\begin{equation}
y_t - y_{t-1}= \omega (x_t - x_{t-1})+ u_t+\rho u_{t-1},
\label{eq:Wrightlawrho}
\end{equation}
where $u_t \sim \mathcal{N}(0,\sigma_u^2)$. We treat $\rho$ as a known parameter. Moreover, we will assume that it is the same for all technologies and we will estimate it as the average of the $\tilde \rho_j$ estimated on each technology separately (as described in the next section). This is a delicate assumption, but it is motivated by the fact that many of our time series are too short to estimate a specific $\rho_j$ reliably, and assuming a universal, known value of $\rho$ allows us to keep analytical tractability.

The forecasts are made exactly as before, but the forecast error now is
\begin{equation}
\mathcal{E}_\tau = \sum_{j=2}^{m+1} H_j [v_j+\rho v_{j-1}]+\sum_{T=t+1}^{t+\tau} [v_T+\rho v_{T-1}],
\end{equation}
where the $H_j$ are defined as
\begin{equation}
H_j=-\frac{\sum_{i=t+1}^{t+\tau} X_{i}}{\sum_{i=2}^{t}X_i^2} X_j.
\end{equation}
The forecast error can be decomposed as a sum of independent normally distributed variables, from which the variance can be computed as
\begin{align}
\begin{split}
E[\mathcal{E}_\tau^2] &= \sigma_u^2 \Big( \rho^2 H_2^2 + \sum_{j=2}^{m} (H_j+\rho H_{j+1})^2 \\
&+(\rho+H_{m+1})^2
+(\tau-1) (1+\rho)^2 +1 \Big).
\label{eq:MSFEIMAX}
\end{split}
\end{align}

When we will do real forecasts (Section \ref{section:PV}), we will take the rate of growth of future cumulative production as constant. If we also assume that the growth rates of past cumulative production were constant, we have $X_i=r$ and thus $H_i=-\tau/m$ for all $i$. As expected from Sahal's identity, simplifying Eq.\eqref{eq:MSFEIMAX} under this assumption gives Eq.\eqref{eq:MSFEMT} where $\theta$ is substituted by $\rho$ and $\sigma_v$ is substituted by $\sigma_u$,
\begin{equation}
E[\mathcal{E}_\tau^2] = \sigma_u^2
\Big[-2\rho + \Big(1+\frac{2(m-1)\rho}{m} +\rho^2\Big) A \Big].
\label{eq:MSFEIMAX2}
\end{equation}
In practice we compute $\hat{\sigma}_\eta$ using Eq. \eqref{eq:sigmahat}, so that $\sigma_u$ may be estimated as
\[
\hat{\sigma}_u=\sqrt{\hat{\sigma}_\eta^2/(1+\rho^2)},
\]
suggesting the normalized error $\mathcal{E}_\tau^2/\hat{\sigma}_\eta^2$. 
To gain more intuition on Eq.\eqref{eq:MSFEIMAX2}, and propose a simple, easy to use  formula, note that for $\tau \gg 1$ and $m \gg 1$ it can be approximated as
\begin{equation}
E\Big[\left(\frac{\mathcal{E}_\tau}{\sigma_\eta}\right)^2 \Big] \approx \frac{(1+\rho)^2}{1+\rho^2}\Big(\tau+\frac{\tau^2}{m} \Big).
\label{eq:MSFEIMAX3}
\end{equation}

For all models (Moore and Wright with and without autocorrelated noise), having determined the variance of the forecast errors we can normalize them so that they follow a Standard Normal distribution
\begin{equation}
\frac{\mathcal{E}_\tau}{\sqrt{E[\mathcal{E}_\tau^2]}} \sim \mathcal{N}(0,1)  
\end{equation}
In what follows we will replace $\sigma_\eta^2$ by its estimated value, so that when $\mathcal{E}$ and $\hat{\sigma}_\eta$ are independent the reference distribution is Student. However, for small sample size $m$ the Student distribution is only a rough approximation, as shown in Appendix \ref{appendix:checktheo}, where it is also shown that the theory works well when the variance is known, or when the variance is estimated but $m$ is large enough.

\subsection{Comparing Moore and Wright at different forecast horizons}
\label{section:compareMooreWright}

One objective of the paper is to compare Moore's law and Wright's law forecasts. To normalize Moore's law forecasts, \citet{farmer2016how} used the estimate of the variance of the delta log cost time series $\hat{K}^2$, as suggested by Eq. \eqref{eq:MSFEmoore}, i.e.
\begin{equation}
\epsilon_{M}=\mathcal{E}_{M}/\hat{K},
\label{eq:moorenormmooreMSFE}
\end{equation}
To compare the two models, we propose that Wright's law forecast errors can be normalized by the very same value
\begin{equation}
\epsilon_{W}=\mathcal{E}_{W}/\hat{K},
\label{eq:moorenormwrightMSFE}
\end{equation}
Using this normalization, we can plot the normalized mean squared errors from Moore's and Wright's models at each forecast horizon. These are directly comparable, because the raw errors are divided by the same value, and these are meaningful because Moore's normalization ensures that the errors from different technologies are comparable and can reasonably be aggregated. In the context of comparing Moore and Wright, when pooling the errors of different forecast horizons we also use the normalization from Moore's model (neglecting autocorrelation for simplicity), $A \equiv \tau+\tau^2/m$ (see \citet{farmer2016how} and Eqs. \ref{eq:MSFEmoore}
and \ref{eq:A}).

\section{Empirical results}
\label{section:section:results}

\subsection{The data}

\begin{figure}[H]
	\includegraphics[height=75mm]{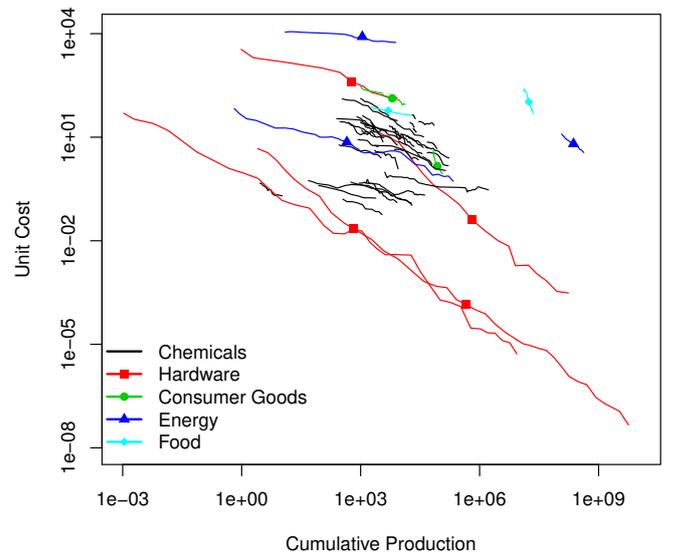}
	\caption{Scatter plot of unit costs against cumulative production.}
	\label{fig:expecurves}
\end{figure}

We mostly use data from the performance curve database\footnote{The data can be accessed at pcdb.santafe.edu.} created at the Santa Fe Institute by Bela Nagy and collaborators from personal communications and from \citet{colpier2002economics,goldemberg2004ethanol,lieberman1984learning,lipman1999experience,zhao1999diffusion,mcdonald2001learning,neij2003experience,moore2006behind,nemet2006beyond} and \citet{schilling2009technology}. We augmented the dataset with data on solar photovoltaics modules taken from public releases of the consulting firms Strategies Unlimited, Navigant and SPV Market Research, which give the average selling price of solar PV modules, and that we corrected for inflation using the US GDP deflator.

\begin{figure}[H]
	\includegraphics[height=75mm]{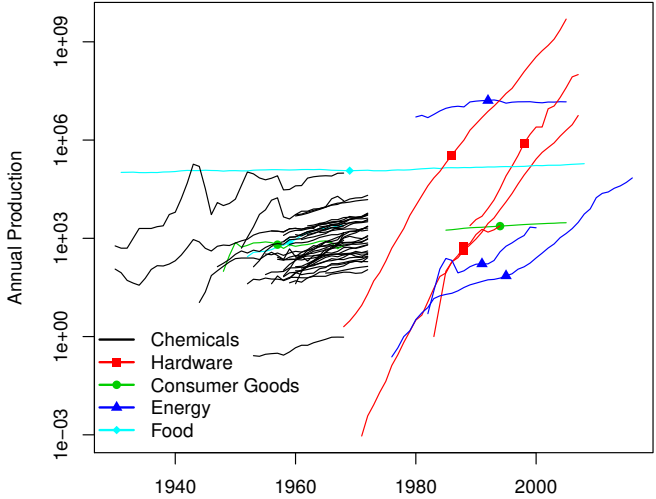}
	\caption{Production time series}
	\label{fig:prod}
\end{figure}

\begin{figure}[H]
	\includegraphics[height=75mm]{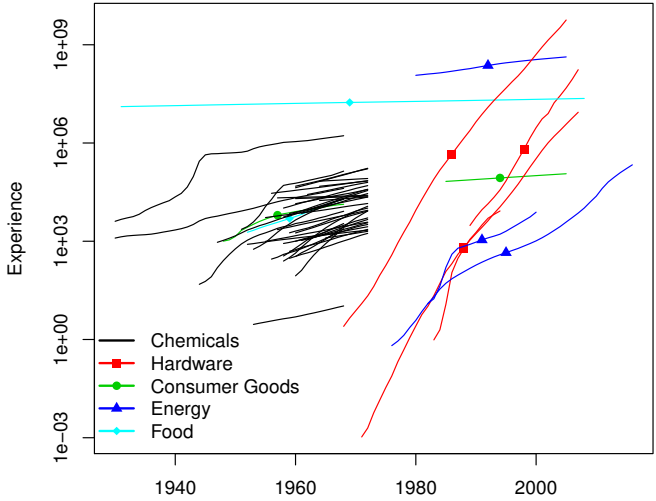}
	\caption{Experience time series}
	\label{fig:expe}
\end{figure}

This database gives a proxy for unit costs\footnote{
In a few cases (milk and automotive), a measure of performance is used instead of costs, and automotive's experience is computed based on distance driven. The main results would not be severely affected by the exclusion of these two time series. Also, unit cost is generally computed from total cost and production of a year or batch, not from actual observation of every unit cost. Different methods may give slightly different results \citep{gallant1968note,womer1983estimation,goldberg2003statistical}, but our dataset is too heterogenous to attempt any correction. Obviously, changes in unit costs do not come from technological progress only, and it is difficult to account for changes in quality, but unit costs are nevertheless a widely used and defensible proxy. }
for a number of technologies over variable periods of time. In principle we would prefer to have data on unit costs, but often these are unavailable and the data is about prices\footnote{This implies a bias whenever prices and costs do not have the same growth rate, as is typically the case when pricing strategies are dynamic and account for learning effects (for instance predatory pricing). For a review of the industrial organization literature on this topic, see \citet{thompson2010learning}.}. Since the database is built from experience curves found in the literature, rather than from a representative sample of products/technologies, there are limits to the external validity of our study but unfortunately we do not know of a database that contains suitably normalized unit costs for all products.

We have selected technologies to minimize correlations between the different time series, by removing technologies that are too similar (e.g. other data on solar photovoltaics). We have also refrained from including very long time series that would represent a disproportionate share of our forecast errors, make the problem of autocorrelation of forecast errors very pronounced, and prevent us from generating many random datasets for reasons of limited computing power. Starting with the set of 53 technologies with a significant improvement rate from \citet{farmer2016how}, we removed DNA sequencing for which no production data was available, and Electric Range which had a zero production growth rate so that we cannot apply the correction to cumulative production described below. We are left with 51 technologies belonging to different sectors (chemical, energy, hardware, consumer durables, and food), although chemicals from the historical reference \citep{BCG} represent a large part of the dataset. For some technologies the number of years is slightly different from \citet{farmer2016how} because we had to remove observations for which data on production was not available.

Figs. \ref{fig:expecurves} shows the experience curves, while Fig. \ref{fig:prod} and \ref{fig:expe} shows production and experience time series, suggesting at least visually that experience time series are ``smoothed'' versions of production time series.

\subsection{Estimating cumulative production}

A potentially serious problem in experience curve studies is that one generally does not observe the complete history of the technology, so that simply summing up observed production misses the experience previously accumulated. There is no perfect solution to this problem. For each technology, we infer the initial cumulative production using a procedure common in the ``R\&D capital'' literature \citep{hall1995exploring}, although not often used in experience curve studies (for an exception see \citet{nordhaus2014perils}). It assumes that production grew as $Q_{t+1}=Q_{t}(1+g_d)$ and experience accumulates as $Z_{t+1}=Z_t+Q_{t}$, so that it can be shown that $Z_{t_0} = Q_{t_0}/g_d$. We estimate the discrete annual growth rate as $\hat{g}_d=\exp( \log(Q_T/Q_{t_0})/(T-1) )-1$, where $Q_{t_0}$ is production during the first available year, and T is the number of available years. We then construct the experience variable as $Z_{t_0} = Q_{t_0}/\hat{g}_d$ for the first year, and $Z_{t+1}=Z_t+Q_{t}$ afterwards.

Note that this formulation implies that experience at time $t$ does not include production of time $t$, so the change in experience from time $t$ to $t+1$ does not include how much is produced during year $t+1$. In this sense we assume that the growth of experience affects technological progress with a certain time lag. We have experimented with slightly different ways of constructing the experience time series, and the aggregated results do not change much, due to the high persistence of production growth rates. 

A more important consequence of this correction is that products with a small production growth rate will have a very important correction for initial cumulative production. In turn, this large correction of initial cumulative production leads to significantly lower values for the annual growth rates of cumulative production. As a result, the experience exponent $\hat \omega$ becomes larger than if there was no correction. This explains why products like milk, which have a very low rate of growth of production, have such a large experience exponent. Depending on the product, this correction may be small or large, and it may be meaningful or not. Here we have decided to use this correction for all products.

\subsection{Descriptive statistics and Sahal's identity}

\begin{figure}[H]
\includegraphics[height=75mm]{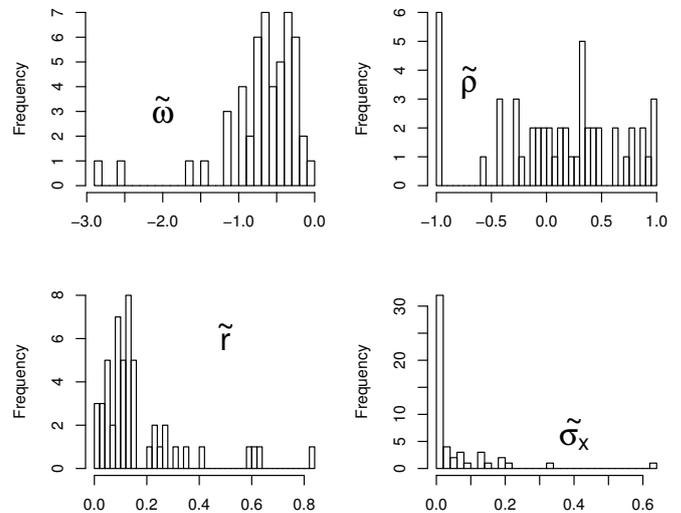}
\caption{Histograms of estimated parameters.}
\label{fig:histparaW}
\end{figure}

\begin{figure}[H]
\includegraphics[height=75mm]{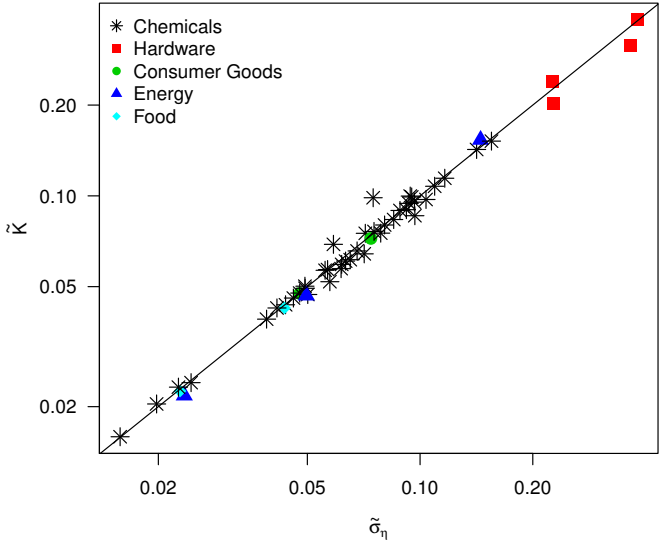}
\caption{Comparison of residuals standard deviation from Moore's and Wright's models.}
\label{fig:Kvigmae}
\end{figure}

Table~\ref{table:parameterestimates} summarizes our dataset, showing in particular the parameters estimated using the full sample.
Fig.~\ref{fig:histparaW} complements the table by showing histograms for the distribution of the most important parameters. Note that we denote estimated parameters using a tilde because we use the full sample (when using the small rolling window, we used the hat notation). To estimate the $\tilde{\rho}_j$, we have used a maximum likelihood estimation of Eq. \ref{eq:Wrightlawrho} (letting $\tilde{\omega}_{MLE}$ differ from $\tilde{\omega}$).
Fig. \ref{fig:Kvigmae}  compares the variance of the noise estimated from Wright's and Moore's models. These key quantities express how much of the change in (log) cost is left unexplained by each model; they also enter as direct factor in the expected mean squared forecast error formulas. The lower the value, the better the fit and the more reliable the forecasts. The figure shows that for each technology the two models give similar values; see Table~\ref{table:parameterestimates}).

\begin{table*}[h!]
\centering
\scalebox{0.85}{
\begin{tabular}{|l|c|cc|cc|cc|ccc|}
  \hline
& & \multicolumn{2}{|c|}{Cost} & \multicolumn{2}{|c|}{Production} & \multicolumn{2}{|c|}{Cumul. Prod.} & \multicolumn{3}{|c|}{Wright's law}\\
& T & $\tilde{\mu}$ &$\tilde{K}$ &$\tilde{g}$ &$\tilde{\sigma}_q$ &$\tilde{r}$ &$\tilde{\sigma}_x$ &$\tilde{\omega}$ &$\tilde{\sigma_{\eta}}$ &$\tilde{\rho}$ \\
\hline
Automotive &   21 & -0.076 & 0.047 & 0.026 & 0.011 & 0.027 & 0.000 & -2.832 & 0.048 & 1.000 \\ 
  Milk &   78 & -0.020 & 0.023 & 0.008 & 0.020 & 0.007 & 0.000 & -2.591 & 0.023 & 0.019 \\ 
  IsopropylAlcohol &    9 & -0.039 & 0.024 & 0.022 & 0.074 & 0.023 & 0.002 & -1.677 & 0.024 & -0.274 \\ 
  Neoprene Rubber &   13 & -0.021 & 0.020 & 0.015 & 0.055 & 0.015 & 0.001 & -1.447 & 0.020 & 0.882 \\ 
  Phthalic Anhydride &   18 & -0.076 & 0.152 & 0.061 & 0.094 & 0.058 & 0.006 & -1.198 & 0.155 & 0.321 \\ 
  TitaniumDioxide &    9 & -0.037 & 0.049 & 0.031 & 0.029 & 0.031 & 0.001 & -1.194 & 0.049 & -0.400 \\ 
  Sodium &   16 & -0.013 & 0.023 & 0.013 & 0.081 & 0.012 & 0.001 & -1.179 & 0.023 & 0.407 \\ 
  Pentaerythritol &   21 & -0.050 & 0.066 & 0.050 & 0.107 & 0.050 & 0.006 & -0.954 & 0.068 & 0.343 \\ 
  Methanol &   16 & -0.082 & 0.143 & 0.097 & 0.081 & 0.091 & 0.004 & -0.924 & 0.142 & 0.289 \\ 
  Hard Disk Drive &   19 & -0.593 & 0.314 & 0.590 & 0.307 & 0.608 & 0.128 & -0.911 & 0.364 & -0.569 \\ 
  Geothermal Electricity &   26 & -0.049 & 0.022 & 0.043 & 0.116 & 0.052 & 0.013 & -0.910 & 0.023 & 0.175 \\ 
  Phenol &   14 & -0.078 & 0.090 & 0.088 & 0.055 & 0.089 & 0.004 & -0.853 & 0.092 & -1.000 \\ 
  Transistor &   38 & -0.498 & 0.240 & 0.585 & 0.157 & 0.582 & 0.122 & -0.849 & 0.226 & -0.143 \\ 
  Formaldehyde &   11 & -0.070 & 0.061 & 0.086 & 0.078 & 0.085 & 0.005 & -0.793 & 0.063 & 0.489 \\ 
  Ethanolamine &   18 & -0.059 & 0.042 & 0.076 & 0.076 & 0.080 & 0.005 & -0.748 & 0.041 & 0.355 \\ 
  Caprolactam &   11 & -0.103 & 0.075 & 0.136 & 0.071 & 0.142 & 0.009 & -0.746 & 0.071 & 0.328 \\ 
  Ammonia &   13 & -0.070 & 0.099 & 0.096 & 0.049 & 0.102 & 0.007 & -0.740 & 0.095 & 1.000 \\ 
  Acrylic Fiber &   13 & -0.100 & 0.057 & 0.127 & 0.126 & 0.137 & 0.020 & -0.726 & 0.056 & -0.141 \\ 
  Ethylene Glycol &   13 & -0.062 & 0.059 & 0.089 & 0.107 & 0.083 & 0.006 & -0.711 & 0.062 & -0.428 \\ 
  DRAM &   37 & -0.446 & 0.383 & 0.626 & 0.253 & 0.634 & 0.185 & -0.680 & 0.380 & 0.116 \\ 
  Benzene &   16 & -0.056 & 0.083 & 0.087 & 0.114 & 0.087 & 0.012 & -0.621 & 0.085 & -0.092 \\ 
  Aniline &   12 & -0.072 & 0.095 & 0.110 & 0.099 & 0.113 & 0.008 & -0.620 & 0.097 & -1.000 \\ 
  VinylAcetate &   13 & -0.082 & 0.061 & 0.131 & 0.080 & 0.129 & 0.010 & -0.617 & 0.065 & 0.341 \\ 
  Vinyl Chloride &   11 & -0.083 & 0.050 & 0.136 & 0.085 & 0.137 & 0.008 & -0.613 & 0.049 & -0.247 \\ 
  Polyethylene LD &   15 & -0.085 & 0.076 & 0.135 & 0.075 & 0.139 & 0.009 & -0.611 & 0.075 & 0.910 \\ 
  Acrylonitrile &   14 & -0.084 & 0.108 & 0.121 & 0.178 & 0.134 & 0.025 & -0.605 & 0.109 & 1.000 \\ 
  Styrene &   15 & -0.068 & 0.047 & 0.112 & 0.089 & 0.113 & 0.008 & -0.585 & 0.050 & 0.759 \\ 
  Maleic Anhydride &   14 & -0.069 & 0.114 & 0.116 & 0.143 & 0.119 & 0.013 & -0.551 & 0.116 & 0.641 \\ 
  Ethylene &   13 & -0.060 & 0.057 & 0.114 & 0.054 & 0.114 & 0.005 & -0.526 & 0.057 & -0.290 \\ 
  Urea &   12 & -0.062 & 0.094 & 0.121 & 0.073 & 0.127 & 0.011 & -0.502 & 0.093 & 0.003 \\ 
  Sorbitol &    8 & -0.032 & 0.046 & 0.067 & 0.025 & 0.067 & 0.002 & -0.473 & 0.046 & -1.000 \\ 
  Polyester Fiber &   13 & -0.121 & 0.100 & 0.261 & 0.132 & 0.267 & 0.034 & -0.466 & 0.094 & -0.294 \\ 
  Bisphenol A &   14 & -0.059 & 0.048 & 0.136 & 0.136 & 0.135 & 0.012 & -0.437 & 0.048 & -0.056 \\ 
  Paraxylene &   11 & -0.103 & 0.097 & 0.259 & 0.326 & 0.228 & 0.054 & -0.417 & 0.104 & -1.000 \\ 
  Polyvinylchloride &   22 & -0.064 & 0.057 & 0.137 & 0.136 & 0.144 & 0.024 & -0.411 & 0.062 & 0.319 \\ 
  Low Density Polyethylene &   16 & -0.103 & 0.064 & 0.213 & 0.164 & 0.237 & 0.069 & -0.400 & 0.071 & 0.473 \\ 
  Sodium Chlorate &   15 & -0.033 & 0.039 & 0.076 & 0.077 & 0.084 & 0.006 & -0.397 & 0.039 & 0.875 \\ 
  TitaniumSponge &   18 & -0.099 & 0.099 & 0.196 & 0.518 & 0.241 & 0.196 & -0.382 & 0.075 & 0.609 \\ 
  Photovoltaics &   41 & -0.121 & 0.153 & 0.315 & 0.202 & 0.318 & 0.133 & -0.380 & 0.145 & -0.019 \\ 
  Monochrome Television &   21 & -0.060 & 0.072 & 0.093 & 0.365 & 0.130 & 0.093 & -0.368 & 0.074 & -0.444 \\ 
  Cyclohexane &   17 & -0.055 & 0.052 & 0.134 & 0.214 & 0.152 & 0.034 & -0.317 & 0.057 & 0.375 \\ 
  Polyethylene HD &   15 & -0.090 & 0.075 & 0.250 & 0.166 & 0.275 & 0.074 & -0.307 & 0.079 & 0.249 \\ 
  CarbonBlack &    9 & -0.013 & 0.016 & 0.046 & 0.051 & 0.046 & 0.002 & -0.277 & 0.016 & -1.000 \\ 
  Laser Diode &   12 & -0.293 & 0.202 & 0.708 & 0.823 & 0.824 & 0.633 & -0.270 & 0.227 & 0.156 \\ 
  Aluminum &   17 & -0.015 & 0.044 & 0.056 & 0.075 & 0.056 & 0.004 & -0.264 & 0.044 & 0.761 \\ 
  Polypropylene &    9 & -0.105 & 0.069 & 0.383 & 0.207 & 0.414 & 0.079 & -0.261 & 0.059 & 0.110 \\ 
  Beer &   17 & -0.036 & 0.042 & 0.137 & 0.091 & 0.146 & 0.016 & -0.235 & 0.043 & -1.000 \\ 
  Primary Aluminum &   39 & -0.022 & 0.080 & 0.088 & 0.256 & 0.092 & 0.040 & -0.206 & 0.080 & 0.443 \\ 
  Polystyrene &   25 & -0.061 & 0.086 & 0.205 & 0.361 & 0.214 & 0.149 & -0.163 & 0.097 & 0.074 \\ 
  Primary Magnesium &   39 & -0.031 & 0.089 & 0.135 & 0.634 & 0.158 & 0.211 & -0.131 & 0.088 & -0.037 \\ 
  Wind Turbine &   19 & -0.038 & 0.047 & 0.336 & 0.570 & 0.357 & 0.337 & -0.071 & 0.050 & 0.750 \\ 
   \hline
 \end{tabular}
} 
\caption{Parameter estimates}
\label{table:parameterestimates}
\end{table*}

\begin{figure}[H]
	\includegraphics[height=75mm]{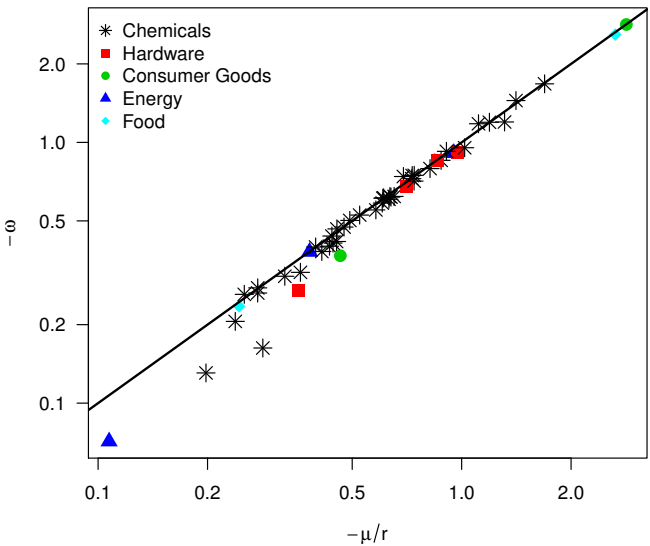}
	\caption{Illustration of Sahal's identity.}
	\label{fig:Sahal}
\end{figure}

Next, we show Sahal's identity as in \citet{nagy2013statistical}. Sahal's observation is that if cumulative production and costs both have exponential trends $r$ and $\mu$, respectively, then costs and production have a power law (constant elasticity) relationship parametrized by $\omega=\mu/r$. One way to check the validity of this relationship is to measure $\mu$, $r$ and $\omega$ independently and plot $\omega$ against $\mu/r$. Fig. \ref{fig:Sahal} shows the results and confirms the relevance of Sahal's identity.

\begin{figure}[H]
	\includegraphics[height=75mm]{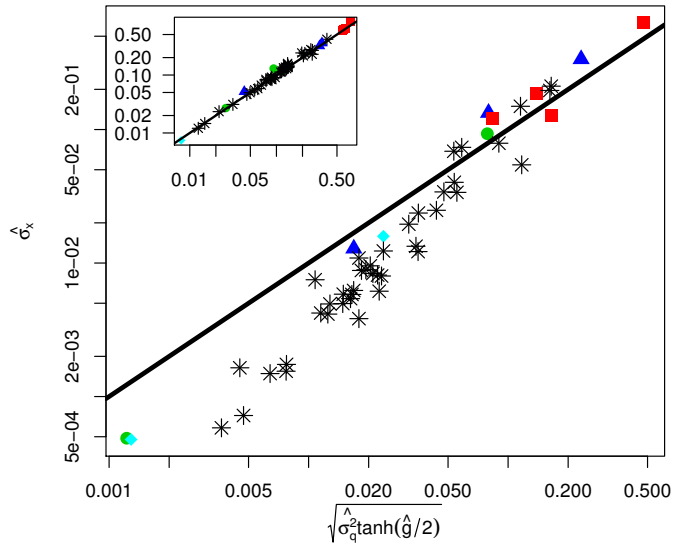}
	\caption{Test of Eq. \ref{eq:sigmaX} relating the volatility of cumulative production to the drift and volatility of production. The inset shows the drift of cumulative production $\hat{r}$ against the drift of production $\hat{g}$.}
	\label{fig:sigmax}
\end{figure}

To explain why Sahal's identity works so well and Moore's and Wright's laws have similar explanatory power, in Section \ref{section:forecasterrors} we have shown that in theory if production grows exponentially, cumulative production grows exponentially with an even lower volatility. Fig. \ref{fig:sigmax} shows how this theoretical result applies to our dataset. Knowing the drift and volatility of the (log) production time series, we are able to predict the drift and volatility of the (log) cumulative production time series fairly well.

\subsection{Comparing Moore's and Wright's law forecast errors}
\label{section:MooreVSWright}

\begin{figure}[H]
\includegraphics[height=75mm]{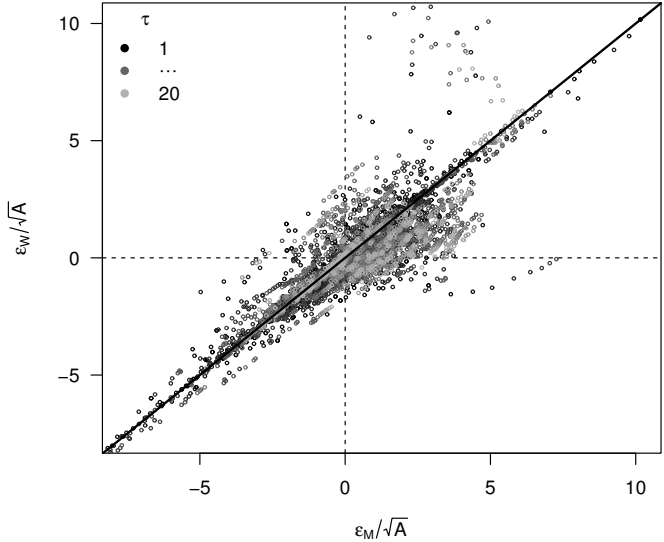}
\caption{Scatter plot of Moore-normalized forecast errors $\epsilon_{M}$ and $\epsilon_{W}$ (forecasts are made using $m=5$). This shows that in the vast majority of cases Wright's and Moore's law forecast errors have the same sign and a similar magnitude, but not always.}
\label{fig:scatterploterror}
\end{figure}

Moore's law forecasts are based only on the cost time series, whereas Wright's law forecasts use information about future experience to predict future costs. Thus, we expect that in principle Wright's forecasts should be better.
We now compare Wright's and Moore's models in a number of ways. The first way is simply to show a scatter plot of the forecast errors from the two models. Fig.~\ref{fig:scatterploterror} shows this scatter plot for the errors normalized by $\hat{K}\sqrt{A}$ (i.e. ``Moore-normalized'', see Eqs \ref{eq:MSFEmoore} \--- \ref{eq:A} and
\ref{eq:moorenormmooreMSFE} and  \ref{eq:moorenormwrightMSFE}), with the identity line as a point of comparison. It is clear that they are highly correlated. When Moore's law over(under)predicts, it is likely that Wright's law over(under)predicts as well, and when Moore's law leads to a large error, it is likely that Wright's law leads to a large error as well. 

\begin{figure}[H]
\includegraphics[height=75mm]{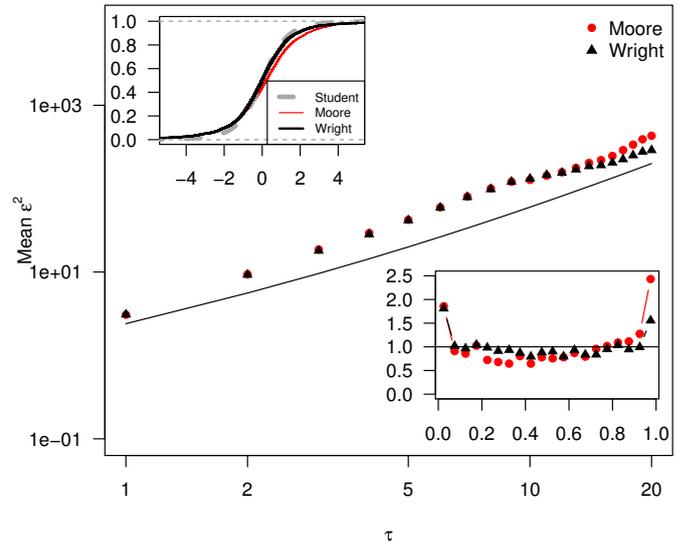}
\caption{Comparison of Moore-normalized forecast errors from Moore's and Wright's models. The main chart shows the mean squared forecast errors at different forecast horizons. The insets show the distribution of the normalized forecast errors as an empirical cumulative distribution function against the Student distribution (top left) and as a probability integral transform against a uniform distribution (bottom right).}
\label{fig:errorgrowth}
\end{figure}

In Fig. \ref{fig:errorgrowth}, the main plot shows the mean squared Moore-normalized forecast error $\epsilon_M^2$ and $\epsilon_W^2$ (see Section \ref{section:compareMooreWright}), where the average is taken over all available forecast errors of a given forecast horizon (note that some technologies are more represented than others). The solid diagonal line is the benchmark for the Moore model without autocorrelation, i.e. the line $y = \frac{m-1}{m-3}A$ \citep{farmer2016how}.
Wright's model appears slightly better at the longest horizons, however there are not many forecasts at these horizons so we do not put too much emphasis on this finding.
The two insets show the distribution of the rescaled Moore-normalized errors $\epsilon/\sqrt{A}$, either as a cumulative distribution function (top left) or using the probability integral transform\footnote{
The Probability Integral Transform is a transformation that allows to compare data against a theoretical distribution by transforming it and comparing it against the Uniform distribution. See for example \citet{diebold1998evaluating}, who used it to construct a test for evaluating density forecasts.
} (bottom right). All three visualizations confirm that Wright's model only slightly outperform Moore's model.

\subsection{Wright's law forecast errors}

\begin{figure}[H]
	\includegraphics[height=75mm]{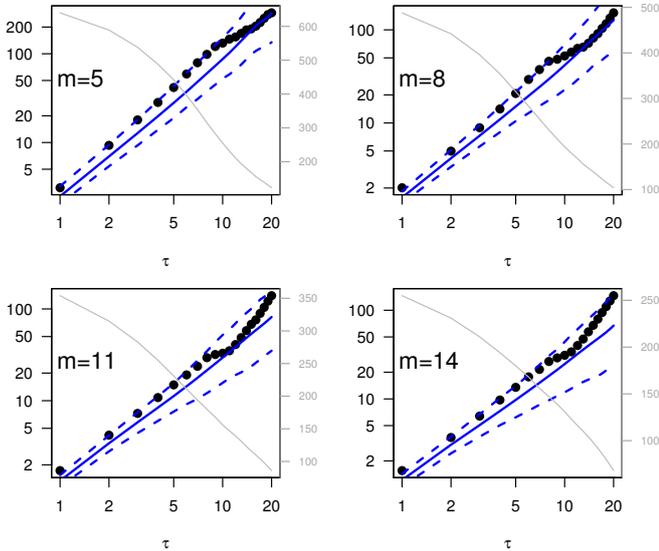}
	\caption{Mean squared Moore-normalized forecast errors of the Wright's law model (Mean $\epsilon^2_{W}$) versus forecast horizon. The 95\% intervals (dashed lines) and the mean (solid line) are computed using simulations as described in the text. The grey line, associated with the right axis, shows the number of forecast errors used to make an average.}
	\label{fig:MonteCarlo}
\end{figure}

In this section, we analyze in detail the forecast errors from Wright's model. We will use the proper normalization derived in Section \ref{section:forecasterrors}, but since it does not allow us to look at horizon specific errors we first look again at the horizon specific Moore-normalized mean squared forecast errors. Fig. \ref{fig:MonteCarlo} shows the results for different values of $m$ (for $m=5$ the empirical errors are the same as in Fig. \ref{fig:errorgrowth}). The confidence intervals are created using the surrogate data procedure described in Section \ref{section:hindcastingandsurrogate}, in which we simulate many random datasets using the autocorrelated Wright's law model (Eq. \ref{eq:Wrightlawrho}) and the parameters of Table \ref{table:parameterestimates} forcing $\rho_j=\rho^*=0.19$ (see below). We then apply the same hindcasting, error normalization and averaging procedure to the surrogate data that we did for the empirical data, and show with blue lines the mean and 95\% confidence intervals. This suggests that the empirical data is compatible with the model (\ref{eq:Wrightlawrho}) in terms of Moore-normalized forecast errors at different horizons.

We now analyze the forecast errors from Wright's model normalized using the (approximate) theory of Section \ref{section:forecasterrors}. Again we use the hindcasting procedure and unless otherwise noted, we use an estimation window of $m=5$ points (i.e. 6 years) and a maximum forecasting horizon $\tau_{max}=20$. To normalize the errors, we need to choose a value of $\rho$. This is a difficult problem, because for simplicity we have assumed that $\rho$ is the same for all technologies, but in reality it probably is not. We have experimented with different methods of choosing $\rho$ based on modelling the forecast errors, for instance by looking for the value of $\rho$ which makes the distribution of normalized errors closest to a Student distribution. While these methods may suggest that $\rho \approx 0.4$, they generally give different values of $\rho$ for different values of $m$ and $\tau_{max}$ (which indicates that some non-stationarity/misspecification is present). Moreover, since the theoretical forecast errors do not exactly follow a Student distribution (see Appendix \ref{appendix:checktheo}) this estimator is biased. For simplicity, we use the average value of $\tilde{\rho}$ in our dataset, after removing the 9 values of $\tilde{\rho}$ whose absolute value was greater than 0.99 (which may indicate a misspecified model). Throughout the paper, we will thus use $\rho^*=0.19$. 

\begin{figure}[H]
	\includegraphics[height=75mm]{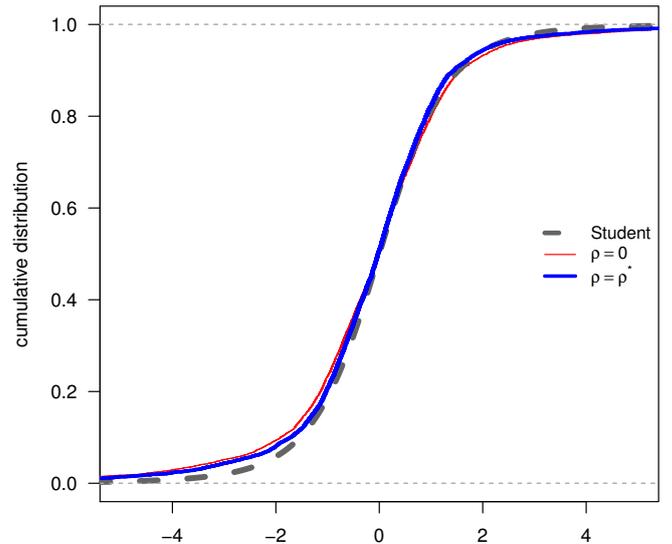}
	\caption{Cumulative distribution of the normalized forecast errors, $m=5$, $\tau_{max}=20$, and the associated $\rho^*=0.19$.}
	\label{fig:densityerror}
\end{figure}

\begin{figure}[H]
	\includegraphics[height=75mm]{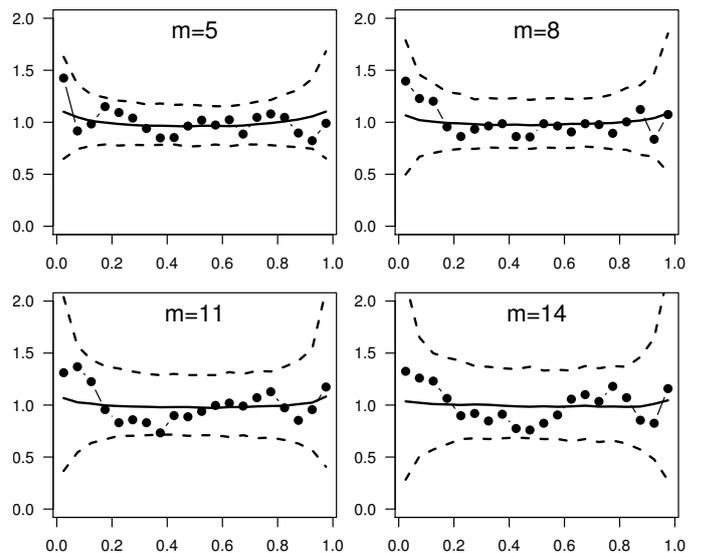}
	\caption{Probability Integral Transform of the normalized forecast errors, for different values of $m$.}
	\label{fig:ChangingmPIT}
\end{figure}

In Fig.~\ref{fig:densityerror}, we show the empirical cumulative distribution function of the normalized errors (for $\rho^*$ and $\rho=0$) and compare it to the Student prediction. In Fig. \ref{fig:ChangingmPIT} we show the probability integral transform of the normalized errors (assuming a Student distribution, and using $\rho=\rho^*$). 
In addition, Fig. \ref{fig:ChangingmPIT} shows the confidence intervals obtained by the surrogate data method, using data simulated under the assumption $\rho=\rho^*$. Again, the results confirm that the empirical forecast errors are compatible with Wright's law, Eq.~(\ref{eq:Wrightlawrho}).

\section{Application to solar photovoltaic modules}
\label{section:PV}

In this section we apply our method to solar photovoltaic modules. Technological progress in solar photovoltaics (PV) is a very prominent example of the use of experience curves.\footnote{
\citet{neij1997use,isoard2001technical,schaeffer2004learning,van2004learning,nemet2006beyond,papineau2006economic,swanson2006vision,alberth2008forecasting,kahouli2009testing,junginger2010technological,candelise2013dynamics}.}
Of course, the limitations of using experience curve models are valid in the context of solar PV modules; we refer the reader to the recent studies by \citet{zheng2014innovation} for a discussion of economies of scale, innovation output and policies; by \citet{de2013predicting} for a discussion of the effects of input prices; and by \citet{hutchby2014moore} for a detailed study of levelized costs (module costs represent only part of the cost of producing solar electricity), and to \citet{farmerUS} for a prediction of levelized solar photovoltaic costs made in 2010.

Historically \citep{wright1936factors,alchian1963reliability}, the estimation of learning curves suggested that costs would drop by 20\% for every doubling of cumulative production, although these early examples are almost surely symptoms of a sample bias. This corresponds to an estimated elasticity of about $\omega=0.33$. As it turns out, estimations for PV have been relatively close to this number. Here we have a progress ratio of $2^{-0.38}=23\%$. A recent study \citep{de2013predicting} contains a more complete review of previous experience curve studies for PV, and finds an average progress ratio of 20.2\%. There are some differences across studies, mostly due to data sources, geographic and temporal dimension, and choice of proxy for experience. Our estimate here differs also because we use a different estimation method (first-differencing), and because we correct for initial cumulative production (most other studies either use available data on initial experience, or do not make a correction; it is quite small here, less than a MW).

In order to compare a forecast based on Wright's law, which gives cost as a function of cumulative production, with a forecast based on Moore's law, which gives cost as a function of time, we need to make an assumption about the production at a given point in time.
Here we provide a distributional forecast for the evolution of PV costs under the assumption that cumulative production will keep growing at exactly the same rate as in the past, but without any variance, and we assume that we know this in advance. One should bear in mind that this assumption is purely to provide a point of comparison.  The two models have a different purpose:  Moore's law gives an unconditional forecast at a given point in time, and Wright's law a forecast conditioned on a given growth rate of cumulative production. While it is perfectly possible to use values from a logistic diffusion model or from expert forecasts, here we illustrate our method based on the exponential assumption to emphasise again the similarity of Moore's and Wright's laws.

Our distributional forecast is
\begin{equation}
y_{T+\tau} \sim \mathcal{N}(\hat{y}_{T+\tau}, V(\hat{y}_{T+\tau}))
\label{eq:forecastEq}
\end{equation}
Following the discussion of Sections \ref{section:forecasterrors} and \ref{section:forecasterrors-Autocorrelation}, the point forecast is
\[
\hat{y}_{T+\tau} = y_T+\tilde{\omega} (x_{T+\tau}-x_T)
\]
and the variance is
\begin{equation}
\begin{split}
V(y_{T+\tau}) &= \frac{\hat{\sigma}_\eta^2}{1+\rho^{*2}} \Big( \rho^{*2} H_2^2 + \sum_{j=2}^{T-1} (H_j+\rho^{*2} H_{j+1})^2 \\
&+(\rho^{*}+H_{T})^2 + (\tau-1) (1+\rho^{*})^2 +1 \Big).
\end{split}
\label{eq:varepsPV}
\end{equation}
where
\[
H_j=-\frac{\sum_{i=T+1}^{T+\tau} X_{i}}{\sum_{i=2}^{T}X_i^2} X_j,
\]
and recalling that $X_t \equiv y_t-y_{t-1}.$

We now assume that the growth of cumulative production is exactly $\tilde{r}$ in the future, so the point forecast simplifies to
\begin{equation}
\hat{y}_{T+\tau} = y_T+\tilde{\omega}\tilde{r}\tau
\label{eq:pfPV}
\end{equation}
Regarding the variance, Eq. \ref{eq:varepsPV} is cumbersome but it can be greatly simplified. First, we assume that past growth rates of experience were constant, that is $X_i=\tilde{r}$ for $i=1\dots T$ (i.e. $\hat{\sigma}^2_x \approx 0$), leading to the equivalent of Eq. \ref{eq:MSFEIMAX2}. As long as production grows exponentially this approximation is likely to be defensible (see Eq. \ref{eq:sigmaX}). Although from Table \ref{table:parameterestimates} we see that solar PV is not a particularly favourable example, we find that this assumption does not affect the variance of forecast errors significantly, at least for short or medium run horizons. 

Since Eq. \ref{eq:MSFEIMAX2} is still a bit complicated, we can further assume that $\tau \gg 1$ and $T\gg 1$, so that we arrive at the equivalent of Eq. \ref{eq:MSFEIMAX3}, that is
\begin{equation}
V(y_{T+\tau}) \approx \hat{\sigma}^2_\eta \frac{(1+\rho^*)^2}{1+\rho^{*2}}\Big(\tau+\frac{\tau^2}{T-1} \Big).
\label{eq:varepsPVsimple}
\end{equation}
This equation is very simple and we will see that it gives results extremely close to Eq. \ref{eq:varepsPV}, so that it can be used in applications.

\begin{figure}[H]
	\includegraphics[height=75mm]{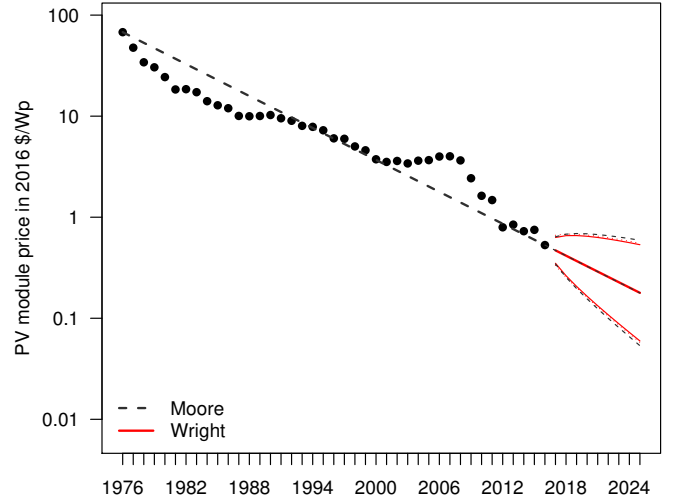}
	\caption{Comparison of Moore's and Wright's law distributional forecasts (95\% prediction intervals).}
	\label{fig:pvMvsW}
\end{figure}

Our point of comparison is the distributional forecast of \citet{farmer2016how} based on Moore's law with autocorrelated noise, and estimating $\theta^*=0.23$ in the same way as $\rho^*$ (average across all technologies of the measured MA(1) coefficient, removing the $|\tilde{\theta}_j| \approx 1$).  All other parameters are taken from Table \ref{table:parameterestimates}. Fig.~\ref{fig:pvMvsW} shows the forecast for the mean log cost and its $95\%$ prediction interval for the two models. The point forecasts of the two models are almost exactly the same because $\hat{\omega}\hat{r}=-0.1209 \approx \hat{\mu}= -0.1213$. Moreover, Wright's law prediction intervals are slightly smaller because $\hat{\sigma}_\eta=0.145<\hat{K}=0.153$. Overall, the forecasts are very similar as shown in Fig. \ref{fig:pvMvsW}. Fig \ref{fig:pvMvsW} does also show the prediction intervals from Eq. \ref{eq:varepsPVsimple}, in red dotted lines, but they are so close to those calculated using Eq. \ref{eq:varepsPV} that the difference can barely be seen.

\begin{figure}[H]
	\includegraphics[height=75mm]{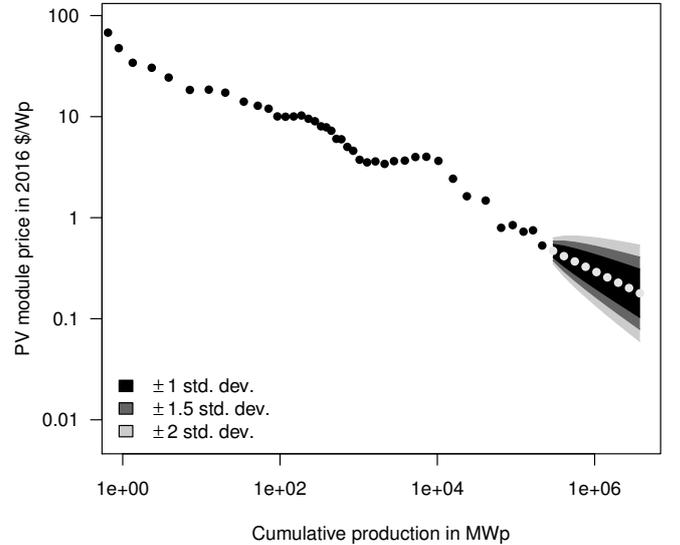}
	\caption{Distributional forecast for the price of PV modules up to 2025, using Eqs. \ref{eq:forecastEq}, \ref{eq:varepsPV} and \ref{eq:pfPV}.}
	\label{fig:pvWexp}
\end{figure}

In Fig. \ref{fig:pvWexp}, we show the Wright's law-based distributional forecast, but against cumulative production. We show the forecast intervals corresponding to 1, 1.5 and 2 standard deviations (corresponding approximately to 68, 87 and 95\% confidence intervals, respectively). The figure also makes clear the large scale deployment assumed by the forecast, with cumulative PV production (log) growth rate of 32\% per year. Again, we note as a caveat that exponential diffusion leads to fairly high numbers as compared to expert opinions \citep{bosetti2012future} and the academic \citep{gan2015quantitative} and professional \citep{masson2013global,IEA2014} literature, which generally assumes that PV deployment will slow down for a number of reasons such as intermittency and energy storage issues. But other studies \citep{zheng2014innovation,jean2015pathways} do take more optimistic assumptions as working hypothesis, and it is outside the scope of this paper to model diffusion explicitly.

\section{Conclusions}

We presented a method to test the accuracy and validity of experience curve forecasts. It leads to a simple method for producing distributional forecasts at different forecast horizons. We compared the experience curve forecasts with those from a univariate time series model (Moore's law of exponential progress), and found that they are fairly similar. This is due to the fact that production tends to grow exponentially\footnote{Recall that we selected technologies with a strictly positive growth rate of production.}, so that cumulative production tends to grow exponentially with low fluctuations, mimicking an exogenous exponential time trend. 
We applied the method to solar photovoltaic modules, showing that if the exponential trend in diffusion continues, they are likely to become very inexpensive in the near future.

There are a number of limitations and caveats that are worth reiterating here: our time series are examples from the literature so that the dataset is likely to have a strong sample bias, which limits the external validity of the results. Also, many time series are quite short, measure technical performance imperfectly, and we had to estimate initial experience in a way that is largely untested. Clearly, the experience curve model also omits important factors such as R\&D. Finally, we make predictions conditional on future experience, which is not the same as doing prediction solely based on time. In settings where production is a decision variable, e.g. \citet{way2017wright}, forecasts conditional on experience are the most useful. However, it remains true that to make an unconditional forecast for a point in time in the future, using Wright's law also requires an additional assumption about the speed of technology diffusion. Thus in a situation of business as usual where experience grows exponentially, using Moore's law is simpler and almost as accurate.

The method we introduce here is closely analogous to that introduced in \citet{farmer2016how}.  Although Moore's law and Wright's law tend to make forecasts of similar quality, it is important to emphasize that when it comes to policy, the difference is potentially very important. While the correlation between costs and cumulative production is well-established, we should stress that the causal relationship is not. But to the extent that Wright's law implies that cumulative production causally influences cost, costs can be driven down by boosting cumulative production. In this case one no longer expects the two methods to make similar predictions, and the method we have introduced here plays a useful role in making it possible to think about not just what the median effect would be, but rather the likelihood of effects of different magnitudes.

\appendix

\section*{Appendix}

\section{Comparison of analytical results to simulations}
\label{appendix:checktheo}

To check whether the analytical theory is reasonable we use the following setting. We simulate 200 technologies for 50 periods. A single time series of cumulative production is generated by assuming that production follows a geometric random walk with drift $g=0.1$ and volatility $\sigma_q=0.1$ (no correction for previous production is made). Cost is generated assuming Wright's law with $\sigma_\eta=0.1$, $\omega=-0.3$ and $\rho=0.6$.

\begin{figure}[H]
	\centering
	\includegraphics[height=75mm]{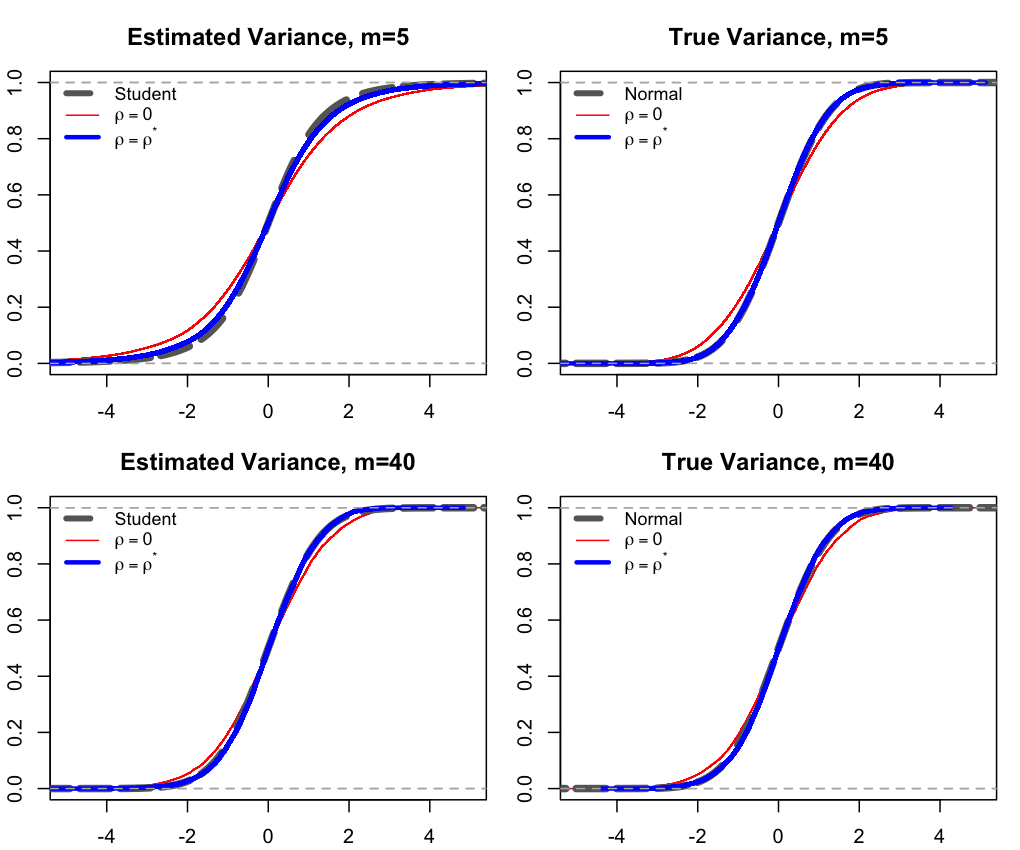}
	\caption{Test of the theory for forecast errors. Top: $m=5$; Bottom: $m=40$. Left: estimated variance; Right: True variance.}
	\label{fig:testtheo}
\end{figure}

Forecast errors are computed by the hindcasting methodology, and normalized using either the true $\rho$ or $\rho=0$. The results are presented in Fig.~\ref{fig:testtheo} for $m=5,40$ and for estimated or true variance ($\hat{\sigma}_v=\hat{\sigma}_\eta/\sqrt{1+\rho^2}$ or $\sigma_v=\sigma_\eta/\sqrt{1+\rho^2}$). In all cases, using the proper normalization factor $\rho=\rho^*$ makes the distribution very close to the predicted distribution (Normal or Student). When $m=5$ and the variance is estimated, we observe a slight departure from the theory as in \citet{farmer2016how}, which seems to be lower for large $m$ or when the true $\sigma_v$ is known.

\begin{figure}[H]
	\centering
	\includegraphics[height=45mm]{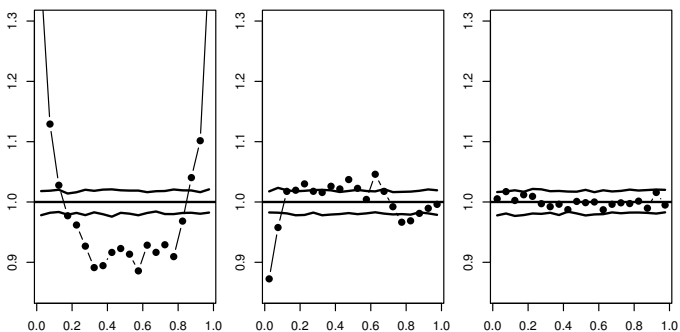}
	\caption{Test of the theory for forecast errors. In the 3 cases $m=5$. On the left, the variance is estimated. In the center, errors are normalized using the true variance. On the right, we also used the true variance but the errors are i.i.d.}
	\label{fig:testtheoPIT}
\end{figure}

To see the deviation from the theory more clearly, we repeat the exercise but this time we apply the probability integral transform to the resulting normalized forecast errors. We use the same parameters, and another realization of the (unique) production time series and of the (200) cost time series. As a point of comparison, we also apply the probability integral transform to randomly generated values from the reference distribution (Student when variance is estimated, Normal when the known variance is used), so that confidence intervals can be plotted. This allows us to see more clearly the departure from the Student distribution when the variance is estimated and $m$ is small (left panel). When the true variance is used (center panel), there is still some departure but it is much smaller. Finally, for the latest panel (right), instead of generating 200 series of 50 periods, we generated 198000 time series of 7 periods, so that we have the same number of forecast errors but they do not suffer from being correlated due to the moving estimation window (only one forecast error per time series is computed). In this case we find that normalized forecast errors and independently drawn normal values are similar.

Overall these simulation results confirm under what conditions our theoretical results are useful (namely, $m$ large enough, or knowing the true variance). For this reason, we have used the surrogate data procedure when testing the errors with small $m$ and estimated variance, and we have used the normal approximation when forecasting solar costs based on almost 40 years of data.

\section{Derivation of the properties of cumulative production}
\label{appendix:sigmaX}

Here we give an approximation for the volatility of the log of cumulative production, assuming that production follows a geometric random walk with drift $g$ and volatility $\sigma_q$. We use saddle point methods to compute the expected value of the log of cumulative production $E[\log Z]$, its variance $\mbox{Var}(\log Z)$ and eventually our quantity of interest $\sigma_x^2 \equiv \mbox{Var}(X) \equiv \mbox{Var}(\Delta \log Z)$. 
The essence of the saddle point method is to approximate the integral by taking into account only that portion of the range of the integration where the integrand assumes large values. More specifically in our calculation we find the maxima of the integrands and approximate fluctuations around these points keeping quadratic and neglecting higher order terms.  Assuming the initial condition $Z(0) = 1$, we can write the cumulative production at time $t$ as $Z_t=1+\sum_{i=1}^te^{g i+\sum_j^i a_j}$, where  $a_1,\ldots ,a_t$  are normally distributed i.i.d. random variables with mean zero and variance $\sigma_q^2$, describing the noise in the production process. $E[\log Z]$ is defined by the (multiple) integral over $a_i$
\begin{eqnarray}
\label{1}
E(\log Z)&=&\int_{-\infty}^{\infty} \log Z\prod_{i=1}^t
\frac{da_i}{\sqrt{2\pi \sigma_q^2}}
\exp\Big[-\frac{a_i^2}{2\sigma_q^2}\Big]\nonumber\\
&=&
\int_{-\infty}^{\infty}e^{S(\{a_i\})}\prod_{i=1}^t
\frac{da_i}{\sqrt{2\pi \sigma_q^2}}\label{DefProbl}
\end{eqnarray}
with $S(\{a_i\})=\log (\log Z)-\sum_{i=1}^t \frac{a_i^2}{2\sigma_q^2}$, which we will
calculate by the saddle point method assuming $\sigma_q^2\ll 1$. 

The saddle point is defined by the system of equations $\partial_i
S(\{a^*_i\})=0, \; \partial_i =\partial/\partial_{a_i}, \; i=1\cdots t$ for
which we can write
\begin{eqnarray}
\label{2}
S(\{a_i\})&=&S(\{a^*_i\})+ \sum_{ij}(a_i-a^*_i)(a_j-a^*_j) G_{ij} \nonumber\\
&&+{\cal O}(\{(a_i-a^*_i)^3\})
\end{eqnarray}
where $a_i^*$ is the solution of the saddle point equations and
$G_{ij}=\frac{1}{2}\partial_i\partial_j S(\{a_i\})\vert_{a_i=a_i^*}$.  In the saddle
point approximation we restrict ourselves to quadratic terms in the expansion
(\ref{2}) which makes the integral (\ref{1}) Gaussian. Then we obtain
\begin{eqnarray}
\label{3}
E(\log Z)=(\det G)^{-1/2}\frac{e^{S(\{a^*_i\})}}{(2\sigma_q^2)^{t/2}} 
\end{eqnarray}
The saddle point equation leads to $\partial_n \left(\log (\log
Z)-\frac{a_n^2}{2\sigma_q^2}\right)=0$, which can be written as
\begin{eqnarray}
\label{4}
a_i&=&\sigma_q^2\,\frac{\partial_iZ}{Z\log Z}=a^*_i+{\cal
  O}(\sigma_q^4)\nonumber\\
a_i^*&=&\sigma_q^2\,\frac{\partial_iZ}{Z\log Z}\Bigg|_{a_i=0}
\end{eqnarray}
Substituting this $a_i^*$ into the $e^{S(\{a^*_i\})}$ term in (\ref{3})
we obtain after some algebra 
\begin{equation}
\label{5}
e^{S(\{a^*_i\})}=\left(\log Z+\frac{\sigma_q^2}{2}
\left.\frac{\sum_{i=1}^t\left(\partial_iZ\right)^2}{Z^2 \log
  Z}\right)\right\vert_{a_i=0} \\
  +O(\sigma_q^4)
\end{equation}
The calculation of $G_{ij}$ as a second derivative gives
\begin{eqnarray}
\label{6}
&&G_{i,j}=\frac{1}{2\sigma_q^2}\Bigg(
\delta_{i,j}+\sigma_q^2\cdot\\
&&\left(\frac{(1+\log Z)\partial_iZ\partial_jZ}{Z^2\log^2Z}-
\frac{\partial_i\partial_jZ}{Z\log Z}
\right)\Bigg\vert_{a_i=0}
\Bigg)+O(\sigma_q^2),\nonumber
\end{eqnarray}
which leads to
\begin{eqnarray}\label{7}
&&(2 \sigma_q^2)^{-t/2}(\det G)^{-1/2}=1+\nonumber\\
&&\left.\frac{\sigma_q^2}{2}
\left(\frac{\sum_{i=1}^t\partial_i^2Z}{Z\log Z}
-\frac{\sum_{i=1}^t(\partial_iZ)^2(1+\log Z)}{Z^2\log^2 Z}\right)
\right\vert_{a_i=0}\nonumber\\
&&+O(\sigma_q^4).
\end{eqnarray}
Here we used the formula $\det G=\exp (tr \log G)$ and an easy expansion of
$\log G$ over $\sigma_q^2$. Now putting formulas (\ref{5}) and (\ref{7}) into
(\ref{3}) we obtain
\begin{eqnarray}
\label{8}
&& E(\log Z)=\log Z\vert_{a_i=0}+\nonumber\\
&&\frac{\sigma_q^2}{2}
\sum_{i=1}^t\left(
\frac{\partial_i^2 Z}{Z}
-\frac{(\partial_i Z)^2}{Z^2}\right)
\Bigg\vert_{a_i=0}+O(\sigma_q^4)
\end{eqnarray}
The calculation of $Z$ and its derivatives at $a_i=0$ is straightforward.  If
$g>0$, for large $t$ it gives the very simple formula
\begin{eqnarray}
E(\log Z(t))|_{t\rightarrow \infty}&=&
g  (t+1)-\log \left(e^{g }-1\right)\nonumber\\
&&+
\frac{\sigma_q^2}{4 \sinh (g )}+O(\sigma_q^4)\label{ElogZ}
\end{eqnarray}

With the same procedure as for (\ref{DefProbl}-\ref{ElogZ}) we calculate the
expectation value of $E(\log^2Z)$, $E(\log Z(t) \log Z(t+1))-E(\log Z(t))E(
\log Z(t+1))$ which leads to similar formulas as (\ref{4}-\ref{8}), but with
different coefficients. The result for $g>0$ and $t\rightarrow \infty $ reads
\begin{eqnarray}
&&\mbox{Var}(\log Z(t))=E(\log^2 Z)-E(\log Z)^2\nonumber\\
&&=\sigma_q^2
\left(\frac{2 e^{g }+1}{1-e^{2 g }}+t\right)
+O(\sigma_q^4)
\end{eqnarray}
and
\begin{eqnarray}
\mbox{Var}(\Delta \log Z)&=&E((\log Z(t+1)-\log Z(t))^2)\nonumber\\
&&-(E(\log Z(t+1)-\log Z(t)))^2\nonumber\\
&= &\sigma_q^2 \tanh \left(\frac{g }{2}\right)+O(\sigma_q^4).
\end{eqnarray}

\bibliographystyle{agsm}
\bibliography{bib-PerfCurves}

\end{document}